\documentclass[3p,10pt]{elsarticle}
\pdfoutput=1

\usepackage{listings}
\usepackage{amssymb}
\usepackage{amsmath}
\usepackage{tikz}

\newcounter{bla}

\begin{document}

\begin{frontmatter}

%% Title, authors and addresses

%% use the tnoteref command within \title for footnotes;
%% use the tnotetext command for the associated footnote;
%% use the fnref command within \author or \address for footnotes;
%% use the fntext command for the associated footnote;
%% use the corref command within \author for corresponding author footnotes;
%% use the cortext command for the associated footnote;
%% use the ead command for the email address,
%% and the form \ead[url] for the home page:
%%
%% \title{Title\tnoteref{label1}}
%% \tnotetext[label1]{}
%% \author{Name\corref{cor1}\fnref{label2}}
%% \ead{email address}
%% \ead[url]{home page}
%% \fntext[label2]{}
%% \cortext[cor1]{}
%% \address{Address\fnref{label3}}
%% \fntext[label3]{}

\title{Finite difference numerical method for the superlattice Boltzmann transport equation and case comparison of CPU(C) and GPU(CUDA) implementations}

%% use optional labels to link authors explicitly to addresses:
%% \author[label1,label2]{<author name>}
%% \address[label1]{<address>}
%% \address[label2]{<address>}

\author{Dmitri Priimak}
\address{Department of Physics, Loughborough University LE11 3TU, United Kingdom}

\begin{abstract}
We present a finite difference numerical algorithm for solving two dimensional spatially homogeneous Boltzmann transport equation which describes electron transport in a semiconductor superlattice
subject to crossed time dependent electric and constant magnetic fields.
The algorithm is implemented both in C language targeted to CPU and in CUDA C language targeted to commodity NVidia GPU. We compare performances and merits of one implementation versus another and discuss various software optimization techniques.
\end{abstract}


\end{frontmatter}
\section{Introduction}
Numerical solutions of Boltzmann Transport Equation (BTE) are of utmost importance in
the modern physics, especially in the research areas of fluid dynamics and semi-classical
description of quantum-mechanical systems. In semiconductors and their nanostructures BTE 
is often used to describe electron dynamics with account of scattering. BTE is often solved 
using Monte-Carlo
method \cite{LorenzoPareschiandGiovanniRusso}. Related to it is the Lattice Boltzmann
Method; it is more recent and very promising \cite{0965-0393-12-6-R01}. Due to
its numerical stability and explicit nature, Lattice Boltzmann
Method lends itself quite well to the
parallel implementations on Graphical Processing Units (GPU) 
\cite{ILBkuCUDAdn,DBLP:journals/corr/MawsonR13,2013arXiv1311.2404J,Kloss20101083}. 
Finite Difference Method (FDM) is the simplest approach to the solution of BTE. However, to attain desirable numerical stability it often requires fully implicit formulation. Recently, a number of new advanced FDMs were developed. In 
\cite{2011CoPhC.182.2445F} a variant of FDM is combined with Monte-Carlo 
method. Fully functional solver for PMOSFET devices, which among other things can utilize 
FDM for solving BTE, was presented in \cite{PMOSFETs}.
Numerical method for spatially non-homogeneous 1D BTE in application to
semiconductor superlattices was recently considered in \cite{Alvaro20124499}.

\begin{figure}
  \centering
  \includegraphics[scale=0.4]{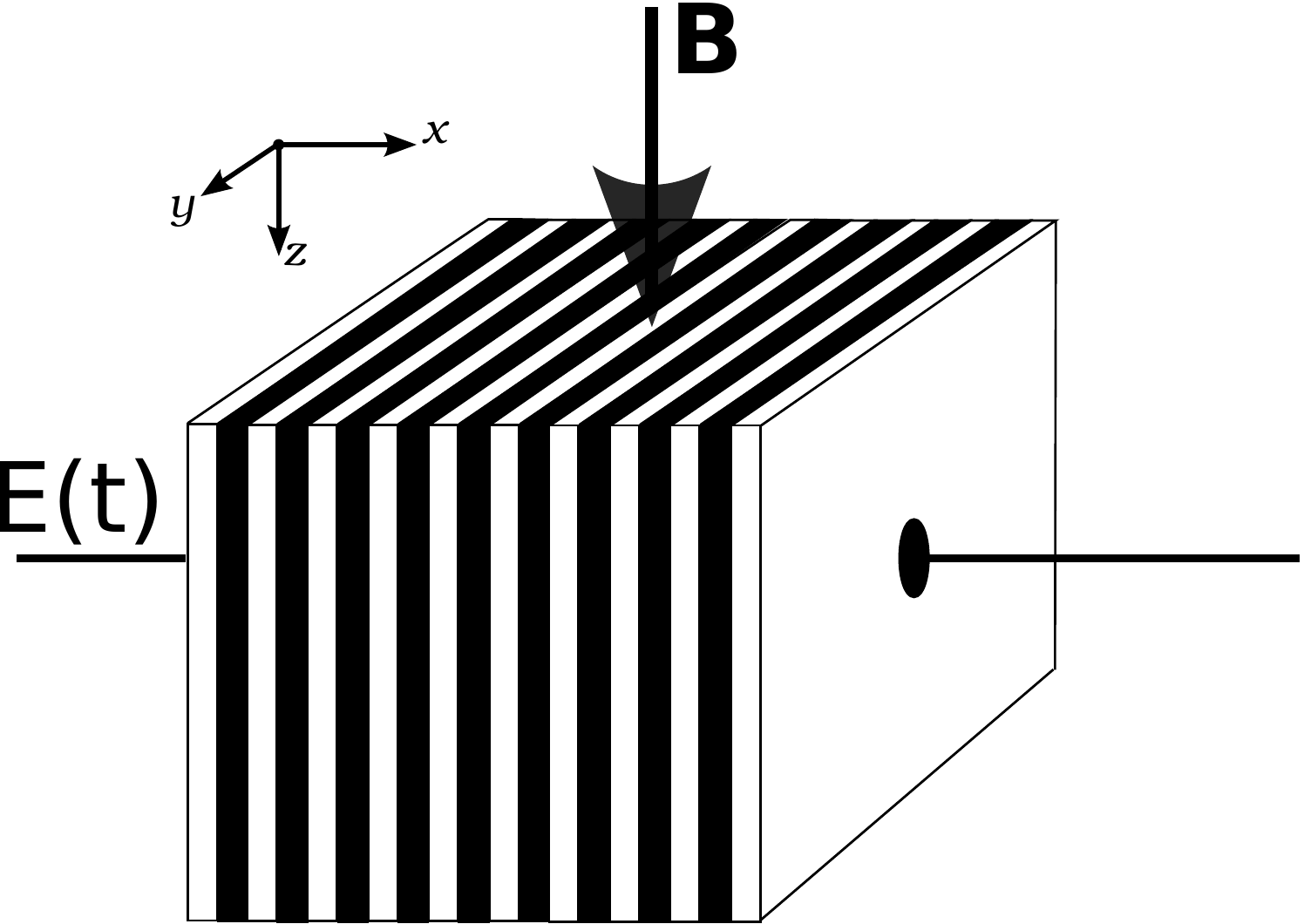}
  \caption{Sketch of a superlattice under the action of time dependent electric $E(t)$ and constant magnetic $\textbf{B}$ fields.
  The electric field is directed along the superlattice axis ($x$-axis) and the perpendicular magnetic field is aligned along
  $z$-axis. Electron motion is considered in $(x,y)$-plane.
  \label{fig:sl_geometry}}
  \end{figure}

In this work, we present a FDM method for solving two-dimensional BTE that describes a semiconductor 
superlattice (SL) subject to a time dependent electric field along the superlattice axis and a constant perpendicular magnetic field. 
Superlattices are artificial periodic 
structures with spatial periods not found in natural solids \cite{Esaki:70}. This relatively large period of SL results in a number of unique physical features, which are interesting not only from the viewpoint of fundamental properties of solids, but also as tools in a realization of promising applications, including the generation and detection of terahertz radiation. Good overview of SL theory and basic experiments can be found in \cite{WAC01,Bass1986237}.
SL and the configuration of applied fields are sketched in Fig. \ref{fig:sl_geometry}. 
In essence this configuration is close to the standard cyclotron resonance
configuration. Terahertz cyclotron resonance in SL has been observed in experiment \cite{PhysRevLett.56.2724}. Especially interesting is the quantity of 
absorption of external ac electric field. When negative it indicates a signal amplification, potentially making possible to consider use of SL as a lasing medium.
Theoretically, this problem was earlier considered in the limiting case of zero temperature \cite{PhysRevLett.103.117401}. This work also indicated that desired signal amplification can occurs within range of parameters where electron distribution is known to be spatially homogeneous. Hence, we also considered electron probability distribution function (PDF) to vary only in the momentum space. 

Our numerical scheme and software that implements it, are used to analyse
electron dynamics in SL at arbitrary temperatures.
It combines Crank-Nicolson 
\cite{CNM} and Leap-Frog algorithms. Leap-Frog algorithm is a variant of 
symplectic 
integrators, which are known to preserve area in the phase space
and are unconditionally stable \cite{0305-4470-39-19-S03,RPITTAAOSI}. We develop 
several implementations
of our numerical scheme. One implementation uses C programming language and is targeted 
for CPU. Other implementations are written in CUDA and are targeted at NVidia 
GPU. Compute Unified Device 
Architecture, also known as CUDA, is parallel computing platform and C/C++ 
language extension for NVidia video cards. 
Different CUDA implementations of our numerical method
primarily highlight various differences in memory access patterns, which are the most common 
bottlenecks for software running on video cards. We verify correctness of 
the method and its implementation by comparing results of our BTE simulations with the available results in the limiting case of zero temperature \cite{PhysRevLett.103.117401,hyarttunable} and with a case when magnetic field is 
absent, for which BTE has exact analytical solution \cite{WAC01}.

\section{Physical model}
  Boltzmann equation governs a time evolution of the electron probability density function (PDF)
  $f(\mathbf{k},\mathbf{r};t)$. For our system it has the following form
    \begin{align}
     \frac{\partial f}{\partial t}+
     \frac{e}{\hbar}\left ( \mathbf{E} + \mathbf{v}\times\mathbf{B} \right )  \frac{\partial f}{\partial\mathbf{k}}&+
     \mathbf{v}(\mathbf{k})\frac{\partial f}{\partial\mathbf{r}} = \left ( \frac{\partial f}{\partial t} \right )_{st} \label{eq:boltzmann} \\
     \mathbf{v}(\mathbf{k})=&\frac{1}{\hbar}\frac{\partial\varepsilon}{\partial\mathbf{k}} \label{eq:v_k}
    \end{align}
where $\mathbf{k}$ is the crystal momentum, $\mathbf{v}(\mathbf{k})$ is the 
electron velocity and $\varepsilon(\mathbf{k})$ is the energy dispersion 
relation for the lowest SL miniband \cite{WAC01}. We make several 
simplifications. As mentioned in the introduction, we assume that $f$ is 
spatially homogeneous and put $\partial f/\partial\mathbf{r}=0$. Secondly, the 
collision
integral $ ( \partial f/\partial t )_{st}$ is taken in the most simplest
form  $(f_0 - f)/\tau$, where $f_0$ is the equilibrium distribution function and 
$\tau$ is the relaxation time constant. We also limit our consideration to electron
transport in a single miniband, which we describe within a tight binding
approximation \cite{Bass1986237}
    \begin{equation}
     \varepsilon=\frac{\hbar^2k^2_y}{2m}-\frac{\Delta_{1}}{2}\cos(k_{x}d)\label{eq:energy_unscaled}
    \end{equation}
    where $\Delta_1$ is the width of the miniband, $d$ is the period of SL and $m$ is 
    the effective electron mass along SL layers.
  To make this system of equations 
  (\ref{eq:boltzmann}) (\ref{eq:v_k}) and (\ref{eq:energy_unscaled}) dimensionless we 
  make the following substitutions.
  \begin{equation}\label{eqs:substitutions}
\begin{alignedat}{2}
\phi_{x}=&k_{x}d &\qquad \phi_{y}=&k_{y}d/\sqrt{\alpha} \\
E/E_{*}\to&E &\qquad E_{*}=&\frac{\hbar}{ed\tau} \\
eB\tau/\sqrt{mm_{x}}\to&B &\qquad t\tau\to &t \\
\alpha=&m/m_{x} &\qquad m_{x}=&\frac{2\hbar^2}{\Delta_{1}d^2} \\
\end{alignedat}
\end{equation}
  And in view of geometry of out system BTE (\ref{eq:boltzmann}) takes form
     \begin{equation}
   	\label{eq:boltzmann_final}
     \frac{\partial f}{\partial t}+
     \left ( E+B\phi_y\right ) \frac{\partial f}{\partial\phi_x}-
     B\sin(\phi_x)\frac{\partial f}{\partial\phi_y}
     = f_0 - f
   \end{equation}
The PDF $f(\phi_x,\phi_y;t)$ can formally extend 
indefinitely along the $y$-axis, but practically this extension is always 
limited by relaxation to the equilibrium distribution 
$f_0(\phi_x,\phi_y)$. In the variables $\phi_x$ and $\phi_y$ normalization 
condition for both $f(\phi_x, \phi_y)$ and $f_{0}(\phi_x, \phi_y)$ takes the following 
form.
	\begin{equation}
		\sqrt{\alpha}\int^{\pi}_{-\pi}\text{d}\phi_x
			\int^{+\infty}_{-\infty}\text{d}\phi_y f(\phi_x,\phi_y)=1 \label{eq:norm_def} 
	\end{equation}
From (\ref{eq:energy_unscaled}) it follows that $f(\phi_x,\phi_y;t)$ is periodic along the $x$-axis 
with the period $2\pi$. Therefore, $\phi_x$ can be considered only within the 
first Brillouin zone defined from $-\pi$ to $\pi$. The periodicity allows us to 
represent both $f$ and $f_0$ as the Fourier series
\begin{align}
f_0 =& \sum_{n=0}^{\infty}a^{(0)}_{n}\cos(n\phi_x) \label{eq:f0_fourier} \\
f = \sum_{n=0}^{\infty}a_{n}&\cos(n\phi_x)+b_{n}\sin(n\phi_x) \label{eq:f_fourier}
\end{align}
where the Fourier coefficients $a^{(0)}_n$, $a_n$ and $b_n$ are functions of $\phi_y$ and the last two are also 
functions of time. In the Fourier representation, BTE (\ref{eq:boltzmann_final}) is transformed to the infinite set of differential equations
\begin{align}
\frac{\partial a_n}{\partial t}=a^{(0)}_n-&a_n - n (E+B\phi_y)b_n + \frac{B}{2} \left (\frac{\partial b_{n+1}}{\partial\phi_y} - \frac{\partial b_{n-1}}{\partial\phi_y} \right ) \label{eq:a_n_dot} \\
\frac{\partial b_n}{\partial t}=-b_n+&n(E+B\phi_y)a_n+\frac{B}{2}\left( \chi(n)\frac{\partial a_{n-1}}{\partial\phi_y} - \frac{\partial a_{n+1}}{\partial\phi_y}\right) \label{eq:b_n_dot} \\
		&\chi(n)= 
		\begin{cases}
	   2 & : n= 1 \\
	   1 & : n\ne 1
	  \end{cases}
\end{align}
	And the normalization condition (\ref{eq:norm_def}) becomes
	\begin{equation}
		2\pi\sqrt{\alpha}\int^{+\infty}_{-\infty}a_0(\phi_{y},t)\text{d}
		\phi_y=1 \label{eq:norm_in_phi}
	\end{equation}
	Eq. (\ref{eq:norm_in_phi}) is later used as one of the tests of accuracy of
	our numerical method. 
The equilibrium PDF $f_0(\phi_x,\phi_y)$ is assumed to be a temperature-dependent Boltzmann distribution, which with 
all normalization constants takes the form
    \begin{align}
	f_0=\frac{1}{2\pi I_0(\mu)}\sqrt{\frac{\mu}{2\pi\alpha}}&\exp{\left \{ \mu\cos(\phi_x)-\frac{\mu}{2}\phi^2_y\right \} } \label{eq:Bolzmann_initial_pdf} \\
	\mu=&\frac{\Delta_1}{2k_{b}T}
	\end{align}
    Where $I_0(\mu)$ is modified Bessel function of zero order. We can use 
    this specific form of $f_0$ to find Fourier coefficients $a^{(0)}_n$ 
	\begin{align}
	a^{(0)}_n=\frac{\sigma(n)I_n(\mu)}{\pi I_0(\mu)}&\sqrt{\frac{\mu}{2\pi\alpha}}\text{exp} \left\lbrace -\frac{\mu}{2} \phi_y^2\right\rbrace \\
	\sigma(n)=&
		\begin{cases}
   		1/2 & : n=0 \\
   		1 & : n\ne 1
  		\end{cases}
	\end{align}
	where $I_n(\mu)$ are the modified Bessel functions of order $n$. 
	
	We also assume that initially at time $t=0$ PDF is in equilibrium state 
	$f(\phi_x, \phi_y; t=0)=f_0(\phi_x,\phi_y)$, i.e. $a_n(t=0)=a^{(0)}_{n}$ and 
	$b_n(t=0)=0$. Equations (\ref{eq:a_n_dot}) and (\ref{eq:b_n_dot}) do not 
	preclude time dependency of both the electric $E$ and magnetic $B$ fields. 
	However, keeping in
	line with the existing research in this field \cite{PhysRevLett.103.117401}, 
	here we consider a magnetic field $B$ to be constant and the electric field 
	to be sum of a constant $E_{dc}$ and monochromatic ac 
	$E_\omega\cos(\omega t)$ components. Thus the total electric field acting on
	electrons in SL is
	$E=E_{dc}+E_{\omega}\cos(\omega t)\label{eq:E_as_a_function_of_time}$.
	We are most interested in the property of absorption of ac electric field, 
	which we defined as
	\begin{equation}
	A=\left\langle \frac{2I_0(\mu)v_{dr}(t)}{I_1(\mu)}\cos(\omega t)\right\rangle_t \label{eq:absorption_def}
	\end{equation}
	Where $v_{dr}(t)$ is the instantaneous
	electron drift velocity along the $x$-axis and $\left\langle\dots\right\rangle_t$ 
	means time averaging over the period of $2\pi/\omega$.
	Negative absorption indicates ac field amplification, paving the way to a lasing 
	medium. To compute absorption we have to let the system to relax to the steady 
	state, which happens over time period of several relaxation time constants 
	$\tau$. In our case, since time is defined in multiples of $\tau$, see 
	(\ref{eqs:substitutions}), in all numerical experiments we let system evolve up 
	to time $t=10$ and then compute averages, such as absorption 
	(\ref{eq:absorption_def}). 
%	Drift velocity used in (\ref{eq:absorption_def}) 
%	follows from (\ref{eq:v_k}) and (\ref{eq:energy_unscaled}) it
%	follows 
%	\begin{equation}
%	v_x=\frac{\Delta_{1}d}{2\hbar}\sin(\phi_x)
%	\end{equation}
	Instantaneous drift velocity along $x$-axis used in (\ref{eq:absorption_def}) is 
	defined as miniband velocity $v_x$ (\ref{eq:v_k}) averaged over PDF 
	\begin{align}
		v_{dr}(t)=&\frac{2d}{\Delta_1\hbar}\iint
			\frac{\partial\varepsilon}{\partial p_x}f(p_x,p_y;t)\text{d}p_x\text{d}p_y \\
			=&\sqrt{\alpha}\int^{\pi}_{-\pi}\text{d}\phi_x\int^{+\infty}_{-\infty}\text{d}\phi_y\sin(\phi_x)f(\phi_x,\phi_y;t)\label{eq:v_dr_generic}
	\end{align}
	which in view of Fourier expansion (\ref{eq:f_fourier}) takes the form
	\begin{equation}
	v_{dr}(t)=\pi\sqrt{\alpha}\int_{-\infty}^{+\infty}b_1(\phi_y;t)\text{d}\phi_y
	\end{equation}
\section{Numerical algorithm}
   	Naive application of method of finite differences
   	to (\ref{eq:a_n_dot}) and (\ref{eq:b_n_dot}) leads to either 
   	unstable and/or computationally intensive numerical system. To 
   	combat this problem we are using several methods at once.
	First, we discretize $a_{n}$ and $b_{n}$ along 
	time and $\phi_y$ axes.
	\begin{equation}
		a^{\textstyle t\leftarrow\text{time step}}_{\textstyle n,m\leftarrow \phi_y \text{lattice step}} \label{eq:a_finite_diffs_definition}
	\end{equation}
	and $n$ is "harmonic number". This forms infinite two-dimensional grid.
	To be computable we limit it to $n\in[0,\dots,N]$ and $m\in[0,\dots,M]$,
	with following boundary conditions.
	\begin{align}
			a^t_{n\notin[0,\dots,N],m\notin [0,\dots,M]}=0 \label{eq:a_boundary}\\
			b^t_{n\notin[1,\dots,N],m\notin [0,\dots,M]}=0 \label{eq:b_boundary}
	\end{align}
	Both upper limits $N$ and $M$ have to be adjusted manually depending on strength
	of electric and magnetic fields and inverse temperature $\mu$, which smears distribution
	function $f$ in the phase space.
	Along the $y$-axis $\phi_y$ is discretized with step 
	$\Delta\phi$ and it becomes function of lattice number $m$.
	
	We write two forms of equations (\ref{eq:a_n_dot}) and (\ref{eq:b_n_dot}).
	One using forward differences and one using partial backward differences, i.e. on 
	the right side of equal sign we are going to write partial derivatives at time $t$ 
	while everything else at time $t+1$ and will follow standard procedure of 
	Crank-Nicolson scheme by adding these two, 
	forward and backward differences equations. First two equations 
	(\ref{eq:a_forward}) and (\ref{eq:b_forward}) below are written in forward
	differencing scheme and last two (\ref{eq:a_backward}) and 
	(\ref{eq:b_backward}) in backward differencing scheme
	\begin{align}
	a^{t+1}_{n,m}-a^{t}_{n,m}=&a^{(0)}_{n,m}\Delta t-a^t_{n,m}\Delta t-
	2b^t_{n,m}\mu^t_{n,m}+B^t_{n,m}\label{eq:a_forward}\\
	b^{t+1}_{n,m}-b^{t}_{n,m}=&-b^t_{n,m}\Delta t+2a^{t}_{n,m}\mu^t_{n,m}+ 
	A^t_{n,m} \label{eq:b_forward} \\
	a^{t+1}_{n,m}-a^{t}_{n,m}=&a^{(0)}_{n,m}\Delta t-a^{t+1}_{n,m}\Delta t-
	2b^{t+1}_{n,m}\mu^{t+1}_{n,m}+B^t_{n,m}\label{eq:a_backward}\\
	b^{t+1}_{n,m}-b^{t}_{n,m}=&-b^{t+1}_{n,m}\Delta t+2a^{t+1}_{n,m}\mu^{t+1}_{n,m}+A^t_{n,m}\label{eq:b_backward}
	\end{align}
	where 
	\begin{align}
	\beta^t_m=&E^t+B^t\phi_y(m) \\
	\mu^t_{n,m}=&n\beta^t_{m}\Delta t/2 \\
	A^t_{n,m}=\frac{\alpha B\Delta t}{4\Delta\phi}(\chi(n)[a^t_{n-1,m+1}-&a^t_{n-1,m-1}]
	-a^t_{n+1,m+1}+a^t_{n+1,m-1}) \\
	B^t_{n,m}=\frac{\alpha B\Delta t}{4\Delta\phi}(b^t_{n+1,m+1}-&b^t_{n+1,m-1}
	-b^t_{n-1,m+1}+b^t_{n-1,m-1})
	\end{align}
	Application of Crank-Nicolson scheme \cite{CNM} leads to
	\begin{align}
	a^{t+1}_{n,m}=\frac{g^t_{n,m}\nu-h^t_{n,m}\mu^{t+1}_{n,m}}{\nu^2+\left(\mu^{t+1}_{n,m}\right)^2}\label{eq:a_t_plus_1}\\
	b^{t+1}_{n,m}=\frac{g^t_{n,m}\mu^{t+1}_{n,m}-h^t_{n,m}\nu}{\nu^2+\left(\mu^{t+1}_{n,m}\right)^2}\label{eq:b_t_plus_1}
	\end{align}
	where 
	\begin{align}
	\nu=&1+\Delta t/2 \\
	\xi=&1-\Delta t/2 \\
	g^t_{n,m}=a^t_{n,m}\xi-
		b^t_{n,m}&\mu^t_{n,m}+B^t_{n,m}+a^{(0)}_{n,m}\Delta t \\
	h^t_{n,m}=b^t_{n,m}\xi+&a^t_{n,m}\mu^t_{n,m}+
		A^t_{n,m}
	\end{align}
	Equation (\ref{eq:a_t_plus_1}, \ref{eq:b_t_plus_1}) can be formally written in the form
	\begin{align}
		\mathbf{z}^{t+1}_{n,m}=&\mathbf{T}(\mathbf{z}^t_{n,m};A^t_{n,m},B^t_{n,m}) \label{eq:time_shift} \\
		\mathbf{z}^{t}_{n,m}=&(a^t_{n,m}, b^t_{n,m})
	\end{align}
	Where $\mathbf{T}$ is an operator that allows us to step from time step $t$ to $t+1$, separated by time interval $\Delta t$. 
	Using this operation as is leads to only conditionally stable numerical
	system, because	$A^t_{n,m}$ and $B^t_{n,m}$ are 
	taken at time $t$, i.e. partially this is still simple forward difference
	scheme.	To combat this problem we introduce two staggered grids 
	$\{\mathbf{z}^0,\mathbf{z}^{1},...\}$ and
	$\{\mathbf{z}^{1/2},\mathbf{z}^{3/2},...\}$, which we call {\em whole} and {\em fractional} one respectively. We then use leap frog method where to
	calculate $\mathbf{z}^{t+1}$ using (\ref{eq:time_shift}) we use $A^{t+1/2}$ and
	$B^{t+1/2}$ computed on fractional grid. Similar operation is performed for
	a step from $t+1/2$ to $t+3/2$. Thus one step from $t$ to $t+1$ becomes two
	steps.
	\begin{align}
	\mathbf{z}^{t+1}_{n,m}=&\mathbf{T}(\mathbf{z}^t_{n,m};A^{t+1/2}_{n,m},B^{t+1/2}_{n,m}) \label{eq:leap_frog_1} \\
	\mathbf{z}^{t+3/2}_{n,m}=&\mathbf{T}(\mathbf{z}^{t+1/2}_{n,m};A^{t+1}_{n,m},B^{t+1}_{n,m}) \label{eq:leap_frog_2}
	\end{align}
	And steps alternate as seen in the following picture
	\begin{center}
	\begin{tikzpicture}[]
	\draw[<-,ultra thick] (1.2,0.0) to [out=0,in=180] (2.3,0.0);
	\draw[<-,ultra thick] (2.7,0.0) to [out=0,in=180] (3.8,0.0);
	\draw[<-,ultra thick] (4.2,0.0) to [out=0,in=180] (5.3,0.0);
	\draw[<-,ultra thick] (5.7,0.0) to [out=0,in=180] (6.8,0.0);
	\draw[<-,ultra thick] (7.2,0.0) to [out=0,in=180] (8.3,0.0);
	\draw[<-,ultra thick] (1.1,0.2) to [out=40,in=140] (3.9,0.2);
	\draw[<-,ultra thick] (2.6,-0.2) to [out=-40,in=-140] (5.4,-0.2);
	\draw[<-,ultra thick] (4.1,0.2) to [out=40,in=140] (6.9,0.2);
	\draw[<-,ultra thick] (5.6,-0.2) to [out=-40,in=-140] (8.4,-0.2);
	\draw[<-,ultra thick] (7.1,0.2) to [out=40,in=160] (8.8,0.56);
	\draw[<-,ultra thick] (0.6,-0.56) to [out=-20,in=-140] (2.4,-0.2);
	\draw[fill] (0.5, -0.5) circle [radius=0.03];
	\draw[fill] (0.4, -0.45) circle [radius=0.03];
	\draw[fill] (8.9, 0.5) circle [radius=0.03];
	\draw[fill] (9, 0.45) circle [radius=0.03];
	\draw[fill] (1, 0) circle [radius=0.1];
	\draw[fill] (2.5, 0) circle [radius=0.1];
	\draw[fill] (4, 0) circle [radius=0.1];
	\draw[fill] (5.5, 0) circle [radius=0.1];
	\draw[fill] (7, 0) circle [radius=0.1];
	\draw[fill] (8.5, 0) circle [radius=0.1];
	\node [right] at (0.5,-0.32) {t+5/2};
	\node [right] at (2.1,0.32) {t+2};
	\node [right] at (3.5,-0.32) {t+3/2};
	\node [right] at (5.1,0.32) {t+1};
	\node [right] at (6.5,-0.32) {t+1/2};
	\node [right] at (8.33,0.32) {t};	
	\end{tikzpicture}
	\end{center}
	This algorithm has to be started first by computing values of $\mathbf{z}^{1/2}$
	using (\ref{eq:time_shift}) with time step $\Delta t/2$. 
    \begin{figure}
   	  \centering
	  \normalsize % use normal font size in the figure
	  \includegraphics[scale=0.43]{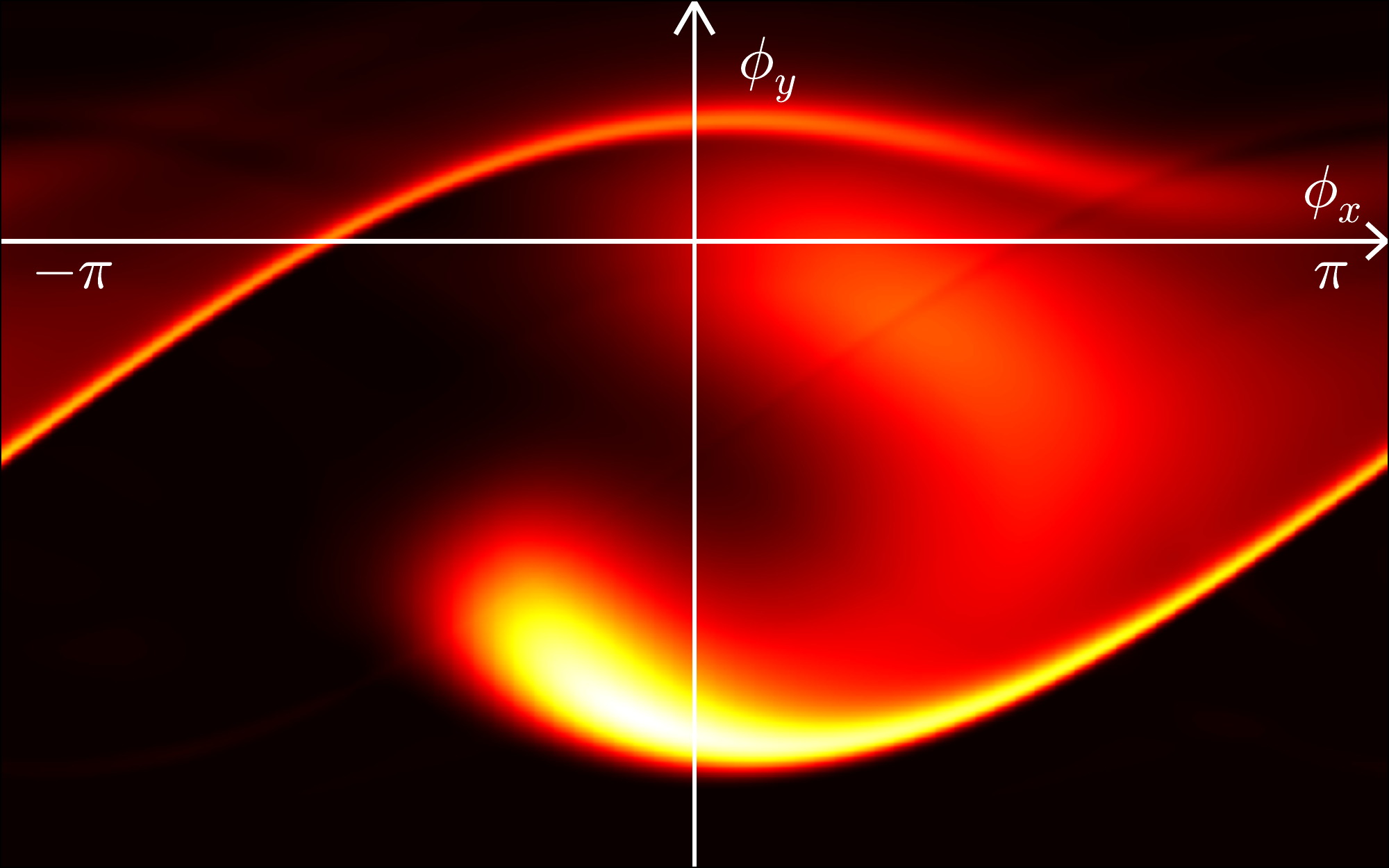}	  
	  \caption{Example of numerical simulation showing a transient
	  behaviour of electron probability density function (PDF) in the phase space
	  $(\phi_x,\phi_y)$ at the time moment $t=1$, 
	  that is before PDF reached its stationary state. Electric and 
	  magnetic fields are applied at time $t=0$.
	  Here the electric field is applied alone the $\phi_x$ axis and the magnetic 
	  field is directed perpendicularly to the plane of the plot. Initially PDF is 
	  concentrated around the center of the plot according to Boltzmann distribution 
	  function (\ref{eq:Bolzmann_initial_pdf}). Other parameters are $E_{dc}=6$ 
	  $B=4$ $E_{\omega}=0$ $\omega=0$ $\mu=3$ $\alpha=0.9496$ $dt=0.0001$
	  \label{fig:tpdf}}
	\end{figure}
	Fig. \ref{fig:tpdf} represents an example of PDF computed by using our method. 
	This heat-map like plot of PDF shows	transient, i.e. before system reaches
	stationary state, response of the system to externally applied, at time $t=0$, 
	electric and 	magnetic fields	according to geometry as shown in Fig. 
	\ref{fig:sl_geometry}.	
\section{Validation of correctness of numerical scheme}
    Complete mathematical analysis of numerical stability and correctness of 
    numerical scheme (\ref{eq:leap_frog_1}) (\ref{eq:leap_frog_2}) is outside 
    of the scope of this paper. 
    However, we can compare solutions obtained by 
    means of our numerical method with solutions obtained by other means for two
    limiting cases: (i) temperature approaches zero and (ii) magnetic field is 
    absent.
    \begin{figure}
    	  \centering
	  \normalsize % use normal font size in the figure
	  \includegraphics[scale=0.75]{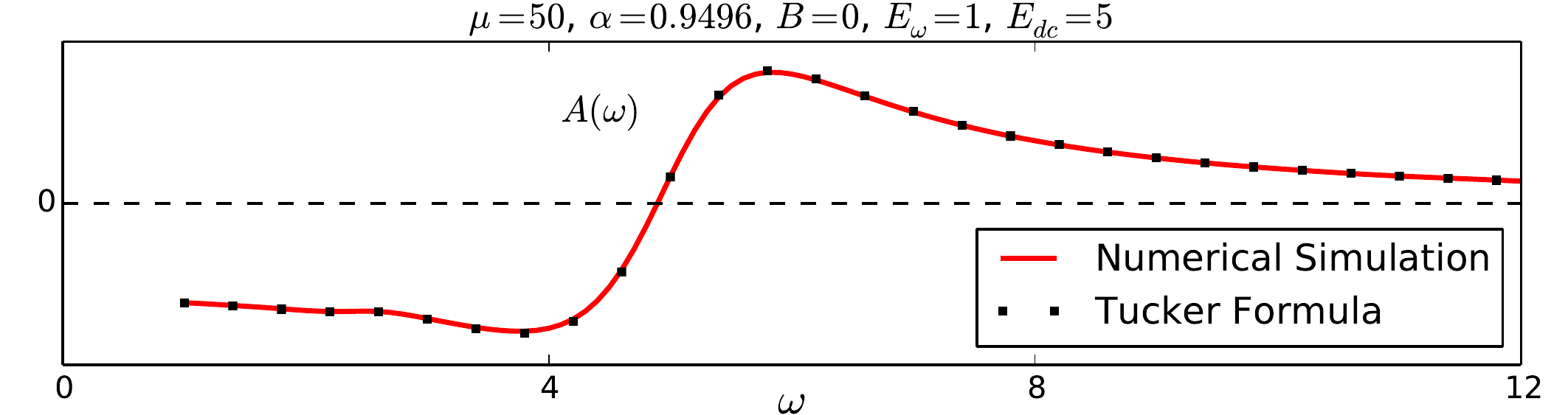}
	  \caption{
	  Absorption $A(\omega)$ for $\mu=50$ $\alpha=0.9496$ $B=0$ $E_{\omega}=1$ 
	  $E_{dc}=5$. Solid red line is computed by means of solving 
	  Boltzmann transport equation using our numerical method. 
	  Black squares are obtained by use of Tucker formula 
	  (\ref{eq:absorption_tacker}). Both solutions match almost perfectly.
	  }.
	  \label{fig:absorption_tacker_cmp}
    \end{figure}   
    If they converge then that validates our approach to solving BTE. 
    Several test runs were performed for different values of external 
    parameters. 
        
    When magnetic field $B$ is zero and temperature is arbitrary, 
	analytical solution to BTE is well known and full analytical expression for
	absorption is known as Tucker formula \cite{WAC01}.
    	\begin{align}
	 A(\omega)=2\frac{I_{1}(\mu)}{I_{0}(\mu)}\sum^{\infty}_{n=-\infty}J_{n}(E_{\omega}/\omega) &
    \left [ J_{n+1}(E_{\omega}/\omega) + J_{n-1}(E_{\omega}/\omega) \right ] 
	    \xi(E_{dc}+n\omega)\label{eq:absorption_tacker} \\
	 \xi(x)=&x/(1+x^2) \label{eq:xi_def}
	\end{align}
	where $J_n(x)$ and $I_n(x)$ are Bessel functions of the first kind and 
	modified Bessel functions, respectively.
    As an example, Fig. \ref{fig:absorption_tacker_cmp} shows results of numerical
    simulation (solid red line) for a given set of parameters 
	and  absorption obtained from analytical Taker formula 
    (\ref{eq:absorption_tacker}), shown in black squares. You 
    can see here that they match nearly perfectly.

    The other limiting case is when temperature goes to zero 
    ($\mu\to\infty$), in which case the equilibrium PDF $f_{0}(\phi_x,\phi_y)$ becomes
    $\delta$-function and instead of BTE (\ref{eq:boltzmann}) we can consider 
    dynamics of a single point in the phase space $\{\phi_x,\phi_y\}$.
	In the absence of dissipation equation (\ref{eq:boltzmann_final}) can be reduced
	to a model of single electron demonstrating pendulum dynamics
	\cite{Bass1986237,PhysRevLett.103.117401,hyarttunable}.
	\begin{align}
	\frac{\text{d}\phi_x}{\text{d}t}=&E+B\phi_y \label{eq:pendulum_phi_x} \\
	\frac{\text{d}\phi_y}{\text{d}t}=&-B\sin(\phi_x) \label{eq:pendulum_phi_y}
	\end{align}
	Which can be trivially solved numerically. Reduction of BTE to the pendulum 
	equation is closely related to the method of characteristic 
	curves \cite{PSSB200844424, 1742-6596-193-1-012004}.
	In the calculation of the drift velocity by means of 
	(\ref{eq:v_k}) (\ref{eq:energy_unscaled}) and 
	(\ref{eq:pendulum_phi_x}) (\ref{eq:pendulum_phi_y}) dissipation can be 
	reintroduced through the use of exponentially decaying function of time as
\begin{equation}
 v_{dr}(t)=\int_{-\infty}^{t}v_x(t) e^{-(t-t_0)} \text{d}t_0  \label{eq:pendulum_drift_velocity}
\end{equation}
	At low temperatures ( high values of $\mu$ ) drift velocity, and 
	therefore absorption (\ref{eq:absorption_def}), computed by use 
	of (\ref{eq:v_dr_generic}) and (\ref{eq:pendulum_drift_velocity}) should
	match. This also serves as a test of correctness of our numerical 
	method and its implementation.
    \begin{figure}
    	  \centering
	  \normalsize % use normal font size in the figure
	  \includegraphics[scale=0.75]{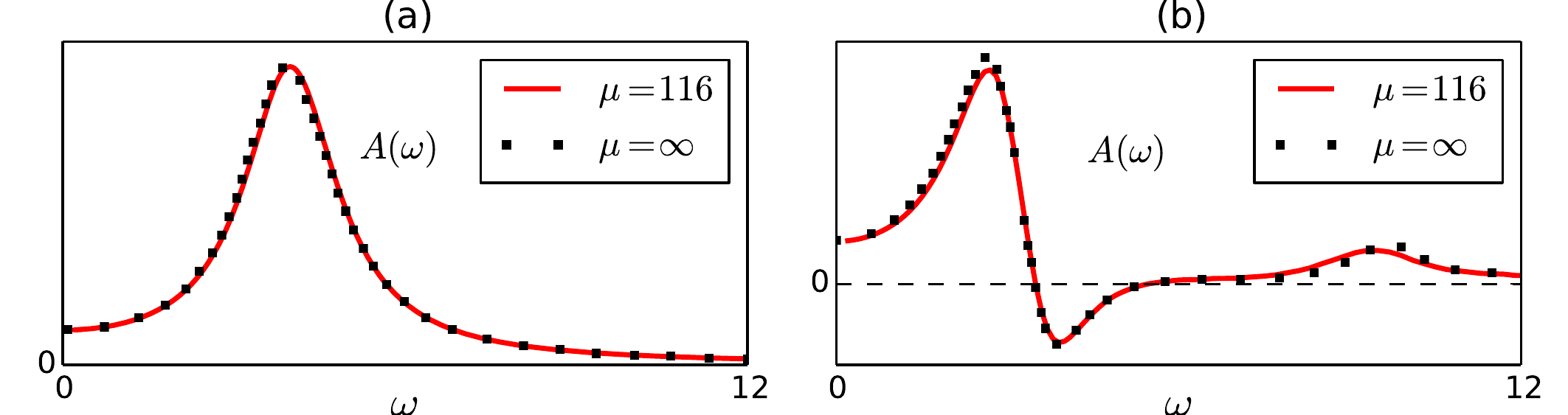}
	  \caption{
	  Absorption $A(\omega)$ for (a) dc electric field $E_{dc}=0$ (Lorentz absorption
	  profile) and (b) $E_{dc}=6$. Other parameters are $\alpha=0.9496$, $B=4$,
	  $E_{\omega}=0.1$. Solid red line is computed by means of solving 
	  Boltzmann transport equation using our numerical method at 
	  relatively low temperate $\mu=116$, sufficiently close to zero. Black squares
	  are computed by solving pendulum equations (\ref{eq:pendulum_phi_x}) 
	  (\ref{eq:pendulum_phi_y}) and (\ref{eq:pendulum_drift_velocity}). They 
	  match very close, but not quite, due to the finite temperature.
	  }
	  \label{fig:asorption_zero_temp_cmp}
    \end{figure}
	In Fig. \ref{fig:asorption_zero_temp_cmp} you can see two cases of 
	comparison of absorption. In the first case absorption is obtained by means of 
	solving BTE 
	using our numerical method (solid red line) at relatively low temperature
	corresponding to $\mu=116$ and in the second case absorption is obtained by means
	of solving pendulum equations (\ref{eq:pendulum_phi_x}), (\ref{eq:pendulum_phi_y})
	and (\ref{eq:pendulum_drift_velocity}) (black squares).
	
\begin{figure}
    	  \centering
	  \normalsize % use normal font size in the figure
	  \includegraphics[scale=0.9]{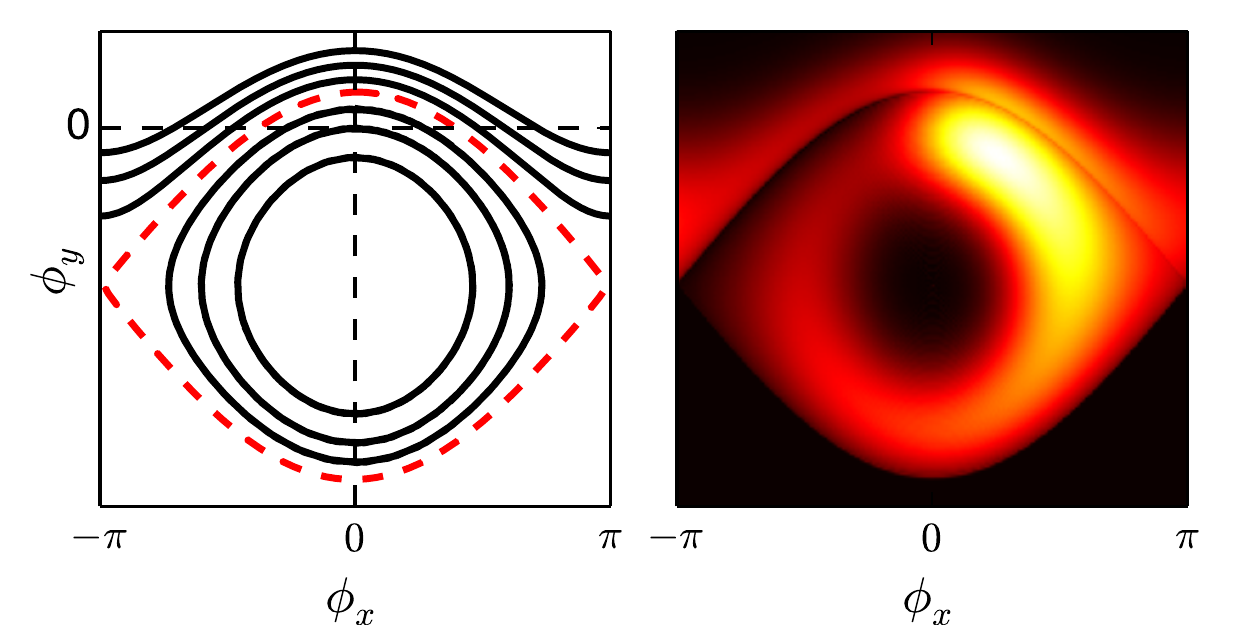}
	  \caption{
	  These two plots show correspondence between phase portrait, 
	  in ($\phi_x$, $\phi_y$) coordinates, of the
	  classical pendulum and final stationary state of electron probability density  
	  function (PDF) for $B=4$ $E_{dc}=6.5$ $E_{\omega}=0$ $\mu=5$. Feature of 
	  separatrix, shown with dashed 
	  red line in the pendulum phase portrait (left panel) can also be seen in 
	  stationary PDF (right panel). Closed trajectories enclosed by separatrix 
	  correspond to cyclotron-like motion. Open trajectories outside of the 
	  separatrix correspond to so-called Bloch oscillations. Note that relation 
	  between proper phase portrait coordinates of classical pendulum 
	  ($\phi_x$, $\dot{\phi}_x$) and the ones used here is defined by the equation 
	  (\ref{eq:pendulum_phi_x}). 
	  }
	  \label{fig:classical_and_Boltzmann_correspondence_map}
\end{figure}

    When both external fields, magnetic and electric ones, are constant 
    ($E_{\omega}=0$) then once PDF reaches stationary state we should see 
    it reflecting characteristic features of classical pendulum.
    In classical pendulum separatrix divides phase space into two regions 
    of closed and open trajectories. Indeed in Fig. 
    \ref{fig:classical_and_Boltzmann_correspondence_map} you can see clear
    correspondence between the phase portrait of classical pendulum on the left and 
    the stationary state of PDF on the right. Notably, this correspondence and 
    especially presence of separatrix appears at arbitrary temperature.
    
    Finally, as mentioned before, norm of PDF at any given 
    moment in time should be equal to one (\ref{eq:norm_def}). Significant 
    deviation of the norm from one can serve as an indicator of instability 
    and/or incorrectness of numerical scheme or improper selection of compute
    parameters, such as too coarse or too fine (due to numerical truncation of 
    float data type) grained mesh or time step. In all of our numerical 
    experiments with compute times up to $t=30$, which is way beyond typical 
    relaxation time, deviation of norm from one was less than $0.01$.

    All these metrics prove that our numerical method is both,
    stable and gives correct solution of BTE (\ref{eq:boltzmann_final}).

\section{CUDA and GPU computing overview}
Modern GPU differ from
CPU in that they have thousands ALUs\footnote{Arithmetic Logic Unit} at the expense
of control hardware and large implicit caches of CPUs. 
    \begin{figure}
   	  \centering
	  \normalsize % use normal font size in the figure
	  \includegraphics[scale=0.6]{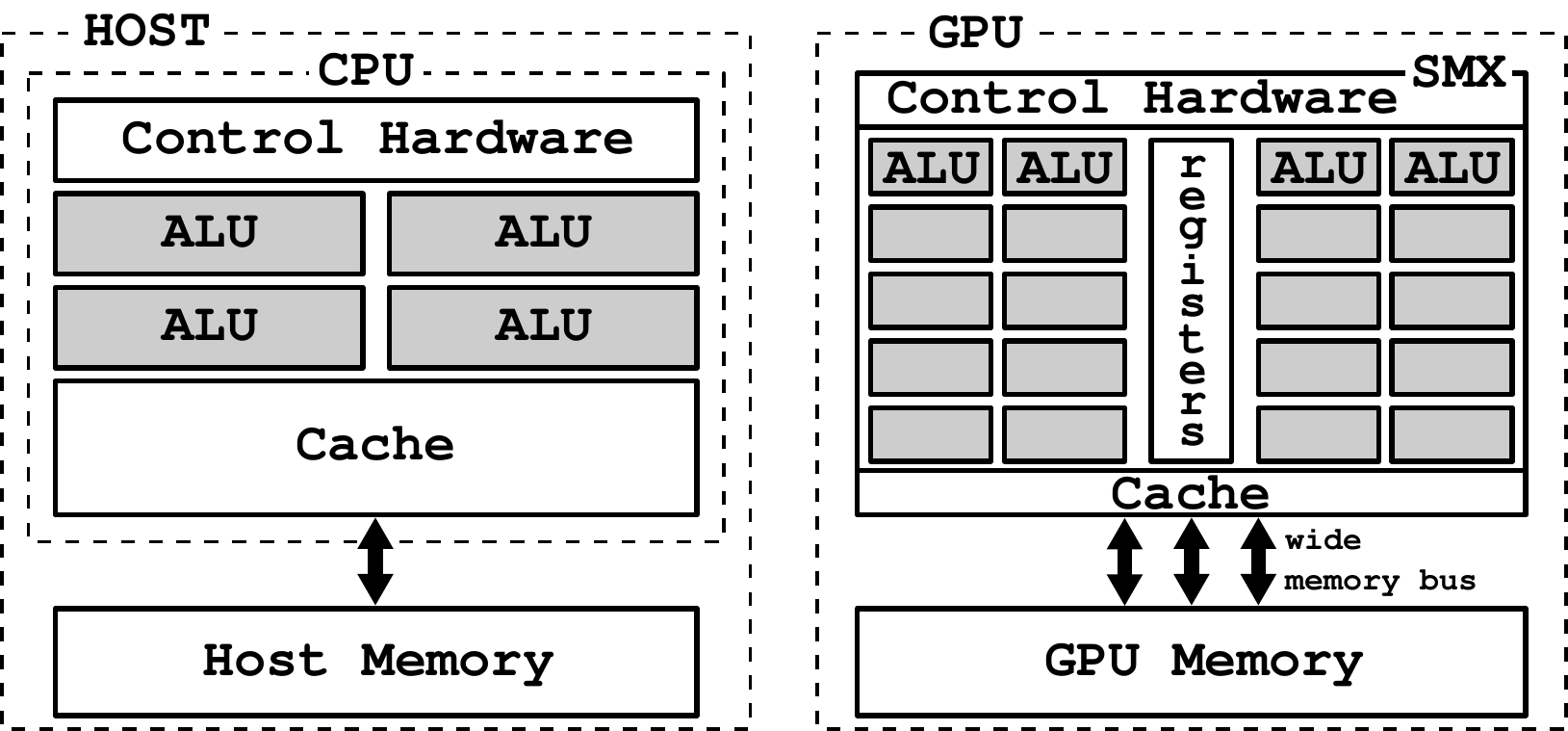}	  
	  \caption{Simplified logical scheme of host computer and GPU that 
	  highlights differences between two. Notable is abundance of ALUs 
	  in modern day GPU, which reaches into thousands and fast wide memory bus 
	  comparing to slow memory bus on the host computer. \label{fig:hcg}}
	\end{figure}
In NVidia video cards these ALUs are known as "CUDA cores". They are grouped into rows and rows into 
larger units with control hardware and explicit caches. These units are known as 
Streaming Multiprocessors (SMX). In turn a single card often contains dozen of SMX 
units. Abundance of ALUs makes for a need of dedicated memory and wide memory bus 
directly on GPU. For example in GTX680 memory bandwidth is approximately 192GB/sec., 
while Intel Core i7 CPU with sandy bridge architecture provides only 37GB/sec 
of aggregate bandwidth. 
In general all of the ALUs in a video card can be executed in parallel, 
although in actuality their execution in SMXs is scheduled in groups of 32
threads known as warps.
This very large degree of parallelism commonly leads to saturation of memory bus 
between on-board GPU memory and SMXs, which means that while programming for GPU
significant speed enhancements can be made by optimising memory access patterns 
and using explicitly available caches \cite{DBLP:journals/corr/MawsonR13,Ryoo:2008:OPA:1345206.1345220,ryoo2008optimization}.
Misaligned and uncoalesced memory access is much slower than indicated by maximum 
available bandwidth. Such access patterns are common problem points 
in CUDA programs. Thus in general CUDA software should try to minimize writing 
and reading to and from memory.
It is common to refer to main computer as {\em host} and installed GPU as 
{\em device}.
Simplified logical layout of host
and GPU can be seen in Fig. \ref{fig:hcg}. Generally speaking GPU can be thought of as an 
explicitly programmable co-processor. White paper describing latest Kepler
architecture of NVidia GPU can be found in \cite{cuda_white_paper}.

CUDA is general purpose computing environment and extension to C and C++ 
languages developed by NVidia. It extends physical abstractions of GPU briefly
described above and 
presents coherent API\footnote{Application Programming Interface} for developing 
general purpose software \cite{cuda_pg}. Basic introduction to CUDA programming 
can 
be found in \cite{Sanders:2010:CEI:1891996} and much more comprehensive one in
\cite{wilt2013cuda}. Software written for a GPU always consist of two parts. One 
part that runs on the host computer, aka host code, and another part that runs
on the device, aka kernel code. It is very common in one program to have several
kernels executing in sequence or in parallel, later one is possible with CUDA
streams.
Kernels are implicitly loaded onto the device by CUDA runtime. They can access
data structures stored on both, on-board device memory and much slower, but
usually much larger host memory. To be placed on the on-board device memory,
data structures have to be created first on the host computer and then
explicitly loaded onto the device (GPU). Execution of a kernel
happens in parallel up to the capacity of the device to do so. Each parallel
flow of execution is known as a {\em thread}.
Threads are organized in hierarchy of grid of blocks of threads. Threads within 
a block can share information through very fast shared memory. Significant amount
of even faster register memory also
available for each thread. Number of blocks and number of threads per block have 
upper limits. For GTX680 GPU grids containing blocks can be three-dimensional
with maximum number of blocks $65535\times 65535\times 65535$ and number of 
threads per blocks is limited to 1024 giving total number of threads an 
astonishing
value of $2^{58}$. We can think of all of them as executing in parallel although
in reality parallelism is ultimately limited by total number of available ALUs. At the simplest
level CUDA programming could be understood as converting inner content of a loop
into kernel code and replacing it with invocation of a kernel on the device.
Inside of a kernel, index variable provided by a loop is replaced by a set 
of implicit variables indicating block number and thread number within a block.
Together with dimensions of a block and grid they can be used to compute an 
equivalent of loop index.
This is illustrated in the code snipped below, where only one of implicit variables
\texttt{threadIdx} is shown. It defines position of thread within a block.
\begin{center}
\begin{tabular}{l c l}
\begin{lstlisting}
for i in [0,...,N]   
   compute(i)
end
\end{lstlisting} &
$\to$ &
\begin{lstlisting}
%*{\color{blue}compute\_kernel}*)()
   i := %*{\bf threadIdx.x}*)
   ...
end
\end{lstlisting}
\end{tabular}
\end{center}
This kernel is later called with parameters indicating number of threads per block
and number of blocks.

\section{C and CUDA implementations}
Using above mentioned algorithm two software packages were written \footnote{Source code for this software is available at https://github.com/priimak/super-lattice-boltzmann-2d}. The C
version that targets CPU and C/CUDA version for running on NVidia GPU.
CUDA version was tested on consumer grade video card GTX680. Both
implementations share the same memory layout.
For C implementation memory layout is not important because computation is limited
by speed of CPU. For CUDA version computation is limited by I/O speed between memory
in a GPU and total number of available ALUs. 
    \begin{figure}
   	  \centering
	  \normalsize % use normal font size in the figure
	  \includegraphics[scale=1]{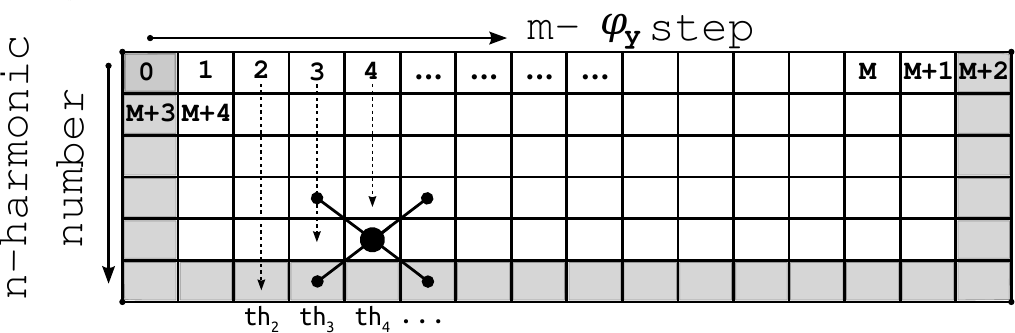}	  
	  \caption{Row-major layout of $a_{n,m}$ and $b_{n,m}$ arrays in linear memory.
	  Grayed out blocks correspond to boundary conditions. They have constant values
	  of 0 and are not modified. This allows to avoid diverging data flows among 
	  groups of threads. One thread is spawned for each $m$ number, shown with dashed 
	  line. Each thread, shown as $th_1$, $th_2$ etc., is responsible for computing next values of $a$ and $b$ arrays
	  across all $n$ harmonic numbers. Cross pattern within array shows
	  neighbouring points that are needed to compute next value of point in the 
	  center (shown with large black circle). \label{fig:array_layout}}
	\end{figure}
Thus layout of arrays storing $a^{(0)}_{n,m}$, $a_{n,m}$ and $b_{n,m}$ 
and memory access patterns makes for the biggest difference in performance.  
We used a row-major layout shown in Fig. \ref{fig:array_layout}. To avoid divergent data flow at the boundaries we shift $m$ index to the right and introduce zero cells
along the perimeter or each array. These zero cells, highlighted in gray in Fig. \ref{fig:array_layout}, ensure that boundary conditions (\ref{eq:a_boundary}) (\ref{eq:b_boundary}) are satisfied without use of \texttt{if} statements. Each thread
computes next values of $a$ and $b$ for all of harmonic numbers for a given $m$ value.
In total we define 9 two-dimensional arrays. $a^{(0)}_{n,m}$ as \texttt{a0(n,m)}.
On the whole grid $a_{n,m}$ as \texttt{a\_h([0,1],n,m)} and $b_{n,m}$ as \texttt{b\_h([0,1],n,m)} and on the fractional one \texttt{a\_f([0,1],n,m)} and
\texttt{b\_f([0,1],n,m)}. First index in \texttt{a} and \texttt{b} arrays can 
take only values of $0$ or $1$ and is used to alternate between current $t$ and
next $t+1$ steps. Both CPU and GPU implementations share this logic and a time
loop, which is shown below. This time loop is a part of a host code.
\begin{center}
\begin{tabular}{l}
Time loop \\
\hline \\
\begin{lstlisting}
cur, nxt := 0, 1
for t in [0, ..., %*T$_{max}$*)]
   compute_time_step(t, cur, nxt)
   cur, nxt := nxt, cur
end
\end{lstlisting} \\
\hline
\end{tabular}
\end{center}
Where \texttt{compute\_time\_step(...)} function performs movement in time 
from $t$ and $t+1/2$ to $t+1$ and $t+3/2$ respectively. Implementation targeted
to CPU implements this function as shown in the following snippet.
\begin{center}
\begin{tabular}{l}
CPU implementation \\
\hline \\
\begin{lstlisting}
for m in [1,...,M+1]
   for n in [0,...,N)
      a_h(nxt, n, m) := 
      	T(a_h(cur,n,m), a_f(cur,n-1,m-1), a_f(cur,n-1,m+1), 
      	  a_f(cur, n+1, m-1), a_f(cur, n+1, m+1))
      b_h(nxt, n, m) := ...
   end
end
\end{lstlisting} \\
\hline
\end{tabular}
\end{center}
where \texttt{T(...)} is implementation of operator (\ref{eq:time_shift}). These
code is repeated once more to compute $a$ and $b$ on fractional grid.
There are several ways this can be transformed into a CUDA code. Two kernels are 
formed. One to move forward in time on the whole grid and another one for 
the fractional grid. Within each of the kernels several variants are possible.
We identify each kernel as $K_x$, where $x$ is implementation number.
The simplest one ($K_1$) is where one thread is allocated for each point of the grid.
In $K_2$ (not shown below) we load $a$ and $b$ into \texttt{\_\_shared\_\_} array, which is first level 
of explicit cache in NVidia GPU. 
\begin{center}
\begin{tabular}{l | l }
$K_1$ & $K_3$ \\
\hline \\
\begin{lstlisting}
kernel_1(...)
   m := ...
   n := ...
   a_h(nxt, n, m) := ...
   b_h(nxt, n, m) := ...
end
\end{lstlisting} &
\begin{lstlisting}
kernel_3(...)
   m := ...
   for n in [0,...,N)
      a_h(nxt, n, m) := ...
      b_h(nxt, n, m) := ...
   end
end
\end{lstlisting} \\
\hline
\end{tabular}
\end{center}
This way we can reduce memory access since nearby threads do access the same
data structures. However, they share very little data and while benefits are
noticeable they are not dramatic. Better and faster code is possible. In $K_3$
kernel shown above, each thread is responsible for computing
$a$ and $b$ for all $n$. In this implementation we do not use shared memory buffer at 
all. Here instruction level parallelism within loop provides very big speed
improvement comparing to kernels $K_1$ and $K_2$. We can unroll loop to gain a
bit more speed. In kernel $K_4$ loops are unrolled twice and in $K_5$ four
times. We can also notice by looking at Fig. \ref{fig:array_layout} that 
steps \texttt{n} and \texttt{n+2} share $a$ and $b$ values at \texttt{n+1}. 
We take advantage of this in kernel $K_5$, where we split each loop into two, each steeping over \texttt{n} with step 2, i.e.
one loop with \texttt{n}=[0,2,4,...] and another one with \texttt{n}=[1,3,5,...].
In each loop we store $a$ and $b$ values at \texttt{n+1} in registers and reuse them.
This provides additional speed boost without any unrolling. Due to register pressure
loop unrolling in $K_6$ does not provide any more speed gain and may even result in 
program becoming slower.

\section{Results}
We compare time needed to perform complete time evolution of 
PDF $f(\phi_x,\phi_y;t)$ up to a given time between all of the above mentioned
implementations, CPU and 6 CUDA implementations. A CPU implementation is a very 
simple single threaded code that does not use any of the 
advanced vector instructions available for modern Intel CPUs. Also 
OpenMP\footnote{OpenMP - Open Multi-Processing, implementation of multithreading} 
version of CPU implementation was tested. The C CPU and host code was compiled by gcc 4.6 with \texttt{-O3} optimization flag. All code
used \texttt{float} data type for storing $a_{n,m}$ and $b_{n,m}$. We found that 
using \texttt{double} data type did not affect precision nor correspondence of 
results
with known solutions. On the other hand GTX680, being consumer grade GPU, lacks in 
efficient capability of performing calculations on \texttt{double} and its 
performance degrades noticeably when switching from \texttt{float} to 
\texttt{double}. Strait 
CPU implementation was single threaded and was tested on "Intel Core i7-3770" 
running at 3.4GHz. CUDA implementations were tested on NVidia GTX680 with 2GB or
RAM. Results
of each test case are presented in Table \ref{tbl:results}. Test cases involved 
running each program 10 times and averaging resulted total run time. Following 
command line parameters were used.
\begin{center}
\begin{tabular}{l}
\begin{lstlisting}
bin/boltzmann_solver display=4 E_dc=7.0 PhiYmin=-6 PhiYmax=6 B=4 t-max=10 E_omega=0.1 \
         omega=10 mu=116 alpha=0.9496 n-harmonics=120 dt=0.0001 g-grid=4000
\end{lstlisting} \\
\end{tabular}
\end{center}
Note that time step \texttt{dt} itself has lower limit due to numerical truncation 
of \texttt{float} data type. We found out that norm of PDF $f(\phi_x,\phi_y)$ starts 
diverging from $1$ for $\text{dt}\le 0.00001$. Run time speed
was compared against CPU implementation baseline and is presented as X times speed up. All 
CUDA implementations were tuned by varying block and grid sizes. Interestingly performance 
of kernel $K_2$, which uses shared memory preloaded with $a^t_{n,m}$ and $b^t_{n,m}$ values 
is only marginally faster then $K_1$. Such behaviour can be explained by the fact that 
values in the shared memory are reused only twice and by presence of computational flow 
divergence around the edges of the cached blocks. It is common to measure
lattice algorithms performance in Million Lattice Updates Per Second (MLUPS). That
parameter is also shown. 
Note that in calculation of MLUPS we count update on 
movement only from time $t$ to $t+1$, i.e. on the whole grid only. 
Otherwise, if
we include updates on the fractional grid, values of MLUPS would have to be doubled.
Attained peak performance is 1094 MLUPS. Faster memory bandwidth and greater number 
of ALUs in later GPUs such as GTX-Titan (memory bandwidth 288GB/sec; 2688 CUDA cores) 
should give significantly higher peak value of MLUPS. For comparison, GTX680 used for
this work has memory bandwidth of 192 GB/sec and 1536 CUDA cores (ALUs).
\begin{table}
\begin{center}
\begin{tabular}{| l | l | l | l | l | l | l | l | l | }
\hline
Impl: & CPU & OpenMP & $K_1$ & $K_2$ & $K_3$ & $K_4$ & $K_5$ & $K_6$ \\ \hline
%Run Time (sec): & $5313\pm$ & $87.1 \pm 0.6$ & $85.3 \pm 0.8$ & $46.8 \pm 0.5$ & $45.3 \pm 0.5$ & $45.5 \pm 0.3$ & $43.85 \pm 0.07$ \\ \hline
Run Time (sec): & $5216$ & 1537 & $87.1$ & $85.3$ & $46.8$ & $45.3$ & $45.5$ & $43.85$ \\ \hline
Speed Up Times: & 1 & 3.4 & 60 & 61 & 111 & 115 & 114 & 118 \\ \hline
MLUPS: & 9 & 31 & 551 & 562 & 1025 & 1059 & 1054 & 1094 \\ \hline
\end{tabular}
\end{center}
\caption{Results\label{tbl:results} of testing of different implementations for 
a given set of parameters. Testing involved simulating evolution of electron 
probability density function up to a given time $t=10$ starting from initial Boltzmann
distribution.
CPU implementation is single threaded running on Intel 
i7-3770 3.4 GHz. OpenMP implementation run on the same CPU with 8 threads. The 
rest are CUDA versions run on GTX680. $K_1$ - one thread per lattice point. $K_2$ 
- same as $K_1$, but using shared memory. In kernel $K_3$ and the rest of kernels 
each thread computes next values for all lattice points with a given $m$-number, 
as seen in Fig. \ref{fig:array_layout}. In $K_4$ main loops are unrolled twice and 
$K_5$ four times. In $K_6$ each loop over $m$ is split in two, each steeping
over 2 elements with lattice values reused in registers. Specific run times
are not so important here, as they depend on the input parameters and GPU card used. Important is relative speed up time measured against baseline CPU implementation.
Last row shows absolute values for MLUPS (Million Lattice Updates Per Second), which 
is a common measure of speed for lattice algorithms. }
\end{table}
On the example of our Boltzmann solver code one can see that even a consumer
grade video card provides significant speed boost to computational tasks
amenable to parallelisation. And if we take in the account low cost of such
video cards, it is now possible to perform computations on the scale which just 
few years ago would require access to the expensive supercomputers. 

\section{Conclusion}
In this work we formulated a numerical method for solving two-dimensional Boltzmann transport 
equation applicable to the semiconductor superlattices. Its correctness and stability were 
verified by comparing results of simulations with results obtained by other means 
in two limiting cases. Several different implementations 
of the algorithm were presented. One written in C for CPU and several for NVidia GPU using 
CUDA. 
We show that even in the most "naive" conversion of C to CUDA 60 fold speed 
improvement is attained. Trying different optimization techniques discussed in this
work CUDA code attains 118 fold speed up over the single threaded C code.

\section{Acknowledgement}
Author expresses his gratitude to Kirill Alekseev and Jukka Isoh\"{a}t\"{a}l\"{a} for very useful discussions on the subject of
this paper and to Timo Hyart for providing some data from his dissertation for comparison with our BTE calculations.

\bibliographystyle{elsarticle-num}
\bibliography{note09}

\begin{thebibliography}{10}
\expandafter\ifx\csname url\endcsname\relax
  \def\url#1{\texttt{#1}}\fi
\expandafter\ifx\csname urlprefix\endcsname\relax\def\urlprefix{URL }\fi
\expandafter\ifx\csname href\endcsname\relax
  \def\href#1#2{#2} \def\path#1{#1}\fi

\bibitem{LorenzoPareschiandGiovanniRusso}
L.~Pareschi, G.~Russo, An introduction to monte carlo method for the boltzmann
  equation, ESAIM: Proceedings 35~(10).

\bibitem{0965-0393-12-6-R01}
D.~Raabe, \href{http://stacks.iop.org/0965-0393/12/i=6/a=R01}{Overview of the
  lattice boltzmann method for nano- and microscale fluid dynamics in materials
  science and engineering}, Modelling and Simulation in Materials Science and
  Engineering 12~(6) (2004) R13.
\newline\urlprefix\url{http://stacks.iop.org/0965-0393/12/i=6/a=R01}

\bibitem{ILBkuCUDAdn}
J.~Tölke, \href{http://dx.doi.org/10.1007/s00791-008-0120-2}{Implementation of
  a lattice boltzmann kernel using the compute unified device architecture
  developed by nvidia}, Computing and Visualization in Science 13~(1) (2010)
  29--39.
\newblock \href {http://dx.doi.org/10.1007/s00791-008-0120-2}
  {\path{doi:10.1007/s00791-008-0120-2}}.
\newline\urlprefix\url{http://dx.doi.org/10.1007/s00791-008-0120-2}

\bibitem{DBLP:journals/corr/MawsonR13}
M.~Mawson, A.~Revell, Memory transfer optimization for a lattice boltzmann
  solver on kepler architecture nvidia gpus, CoRR abs/1309.1983.

\bibitem{2013arXiv1311.2404J}
M.~{Januszewski}, M.~{Kostur}, {Sailfish: a flexible multi-GPU implementation
  of the lattice Boltzmann method}, ArXiv e-prints\href
  {http://arxiv.org/abs/1311.2404} {\path{arXiv:1311.2404}}.

\bibitem{Kloss20101083}
Y.~Kloss, P.~Shuvalov, F.~Tcheremissine,
  \href{http://www.sciencedirect.com/science/article/pii/S1877050910001213}{So%
lving boltzmann equation on \{GPU\}}, Procedia Computer Science 1~(1) (2010)
  1083 -- 1091, <ce:title>ICCS 2010</ce:title>.
\newblock \href
  {http://dx.doi.org/http://dx.doi.org/10.1016/j.procs.2010.04.120}
  {\path{doi:http://dx.doi.org/10.1016/j.procs.2010.04.120}}.
\newline\urlprefix\url{http://www.sciencedirect.com/science/article/pii/S18770%
50910001213}

\bibitem{2011CoPhC.182.2445F}
A.~{Frezzotti}, G.~P. {Ghiroldi}, L.~{Gibelli}, {Solving the Boltzmann equation
  on GPUs}, Computer Physics Communications 182 (2011) 2445--2453.
\newblock \href {http://arxiv.org/abs/1005.5405} {\path{arXiv:1005.5405}},
  \href {http://dx.doi.org/10.1016/j.cpc.2011.07.002}
  {\path{doi:10.1016/j.cpc.2011.07.002}}.

\bibitem{PMOSFETs}
A.-T. Pham, C.~Jungemann, B.~Meinerzhagen,
  \href{http://dx.doi.org/10.1007/s10825-009-0301-3}{On the numerical aspects
  of deterministic multisubband device simulations for strained double gate
  pmosfets}, Journal of Computational Electronics 8~(3-4) (2009) 242--266.
\newblock \href {http://dx.doi.org/10.1007/s10825-009-0301-3}
  {\path{doi:10.1007/s10825-009-0301-3}}.
\newline\urlprefix\url{http://dx.doi.org/10.1007/s10825-009-0301-3}

\bibitem{Alvaro20124499}
M.~\'Alvaro, M.~Carretero, L.~Bonilla,
  \href{http://www.sciencedirect.com/science/article/pii/S0021999112001337}{Nu%
merical method for hydrodynamic modulation equations describing bloch
  oscillations in semiconductor superlattices}, Journal of Computational
  Physics 231~(13) (2012) 4499 -- 4514.
\newblock \href {http://dx.doi.org/http://dx.doi.org/10.1016/j.jcp.2012.02.024}
  {\path{doi:http://dx.doi.org/10.1016/j.jcp.2012.02.024}}.
\newline\urlprefix\url{http://www.sciencedirect.com/science/article/pii/S00219%
99112001337}

\bibitem{Esaki:70}
L.~Esaki, R.~Tsu, \href{http://dx.doi.org/10.1147/rd.141.0061}{Superlattice and
  negative differential conductivity in semiconductors}, IBM J. Res. Dev.
  14~(1) (1970) 61--65.
\newblock \href {http://dx.doi.org/10.1147/rd.141.0061}
  {\path{doi:10.1147/rd.141.0061}}.
\newline\urlprefix\url{http://dx.doi.org/10.1147/rd.141.0061}

\bibitem{WAC01}
A.~Wacker, Semiconductor superlattices: A model system for nonlinear transport,
  Phys.~Rep. 357 (2002) 1.

\bibitem{Bass1986237}
F.~Bass, A.~Tetervov,
  \href{http://www.sciencedirect.com/science/article/pii/0370157386900839}{Hig%
h-frequency phenomena in semiconductor superlattices}, Physics Reports 140~(5)
  (1986) 237 -- 322.
\newblock \href
  {http://dx.doi.org/http://dx.doi.org/10.1016/0370-1573(86)90083-9}
  {\path{doi:http://dx.doi.org/10.1016/0370-1573(86)90083-9}}.
\newline\urlprefix\url{http://www.sciencedirect.com/science/article/pii/037015%
7386900839}

\bibitem{PhysRevLett.56.2724}
T.~Duffield, R.~Bhat, M.~Koza, F.~DeRosa, D.~M. Hwang, P.~Grabbe, S.~J. Allen,
  \href{http://link.aps.org/doi/10.1103/PhysRevLett.56.2724}{Electron mass
  tunneling along the growth direction of (al,ga) as/gaas semiconductor
  superlattices}, Phys. Rev. Lett. 56 (1986) 2724--2727.
\newblock \href {http://dx.doi.org/10.1103/PhysRevLett.56.2724}
  {\path{doi:10.1103/PhysRevLett.56.2724}}.
\newline\urlprefix\url{http://link.aps.org/doi/10.1103/PhysRevLett.56.2724}

\bibitem{PhysRevLett.103.117401}
T.~Hyart, J.~Mattas, K.~N. Alekseev,
  \href{http://link.aps.org/doi/10.1103/PhysRevLett.103.117401}{Model of the
  influence of an external magnetic field on the gain of terahertz radiation
  from semiconductor superlattices}, Phys. Rev. Lett. 103 (2009) 117401.
\newblock \href {http://dx.doi.org/10.1103/PhysRevLett.103.117401}
  {\path{doi:10.1103/PhysRevLett.103.117401}}.
\newline\urlprefix\url{http://link.aps.org/doi/10.1103/PhysRevLett.103.117401}

\bibitem{CNM}
J.~Crank, P.~Nicolson, \href{http://dx.doi.org/10.1007/BF02127704}{A practical
  method for numerical evaluation of solutions of partial differential
  equations of the heat-conduction type}, Advances in Computational Mathematics
  6~(1) (1996) 207--226.
\newblock \href {http://dx.doi.org/10.1007/BF02127704}
  {\path{doi:10.1007/BF02127704}}.
\newline\urlprefix\url{http://dx.doi.org/10.1007/BF02127704}

\bibitem{0305-4470-39-19-S03}
E.~Forest, \href{http://stacks.iop.org/0305-4470/39/i=19/a=S03}{Geometric
  integration for particle accelerators}, Journal of Physics A: Mathematical
  and General 39~(19) (2006) 5321.
\newline\urlprefix\url{http://stacks.iop.org/0305-4470/39/i=19/a=S03}

\bibitem{RPITTAAOSI}
H.~Yoshida, \href{http://dx.doi.org/10.1007/BF00699717}{Recent progress in the
  theory and application of symplectic integrators}, Celestial Mechanics and
  Dynamical Astronomy 56~(1-2) (1993) 27--43.
\newblock \href {http://dx.doi.org/10.1007/BF00699717}
  {\path{doi:10.1007/BF00699717}}.
\newline\urlprefix\url{http://dx.doi.org/10.1007/BF00699717}

\bibitem{hyarttunable}
T.~Hyart, Tunable superlattice amplifiers based on dynamics of miniband
  electrons in electric and magnetic fields.

\bibitem{PSSB200844424}
F.~Brosens, W.~Magnus, \href{http://dx.doi.org/10.1002/pssb.200844424}{Carrier
  transport in nanodevices: revisiting the boltzmann and wigner distribution
  functions}, physica status solidi (b) 246~(7) (2009) 1656--1661.
\newblock \href {http://dx.doi.org/10.1002/pssb.200844424}
  {\path{doi:10.1002/pssb.200844424}}.
\newline\urlprefix\url{http://dx.doi.org/10.1002/pssb.200844424}

\bibitem{1742-6596-193-1-012004}
W.~Magnus, F.~Brosens, B.~Sorée,
  \href{http://stacks.iop.org/1742-6596/193/i=1/a=012004}{Time dependent
  transport in 1d micro- and nanostructures: Solving the boltzmann and
  wigner–boltzmann equations}, Journal of Physics: Conference Series 193~(1)
  (2009) 012004.
\newline\urlprefix\url{http://stacks.iop.org/1742-6596/193/i=1/a=012004}

\bibitem{Ryoo:2008:OPA:1345206.1345220}
S.~Ryoo, C.~I. Rodrigues, S.~S. Baghsorkhi, S.~S. Stone, D.~B. Kirk, W.-m.~W.
  Hwu, \href{http://doi.acm.org/10.1145/1345206.1345220}{Optimization
  principles and application performance evaluation of a multithreaded gpu
  using cuda}, in: Proceedings of the 13th ACM SIGPLAN Symposium on Principles
  and Practice of Parallel Programming, PPoPP '08, ACM, New York, NY, USA,
  2008, pp. 73--82.
\newblock \href {http://dx.doi.org/10.1145/1345206.1345220}
  {\path{doi:10.1145/1345206.1345220}}.
\newline\urlprefix\url{http://doi.acm.org/10.1145/1345206.1345220}

\bibitem{ryoo2008optimization}
S.~Ryoo, C.~I. Rodrigues, S.~S. Baghsorkhi, S.~S. Stone, D.~B. Kirk, W.-m.~W.
  Hwu, Optimization principles and application performance evaluation of a
  multithreaded gpu using cuda, in: Proceedings of the 13th ACM SIGPLAN
  Symposium on Principles and practice of parallel programming, ACM, 2008, pp.
  73--82.

\bibitem{cuda_white_paper}
NVidia,
  \href{http://www.nvidia.com/content/PDF/kepler/NVIDIA-Kepler-GK110-Architect%
ure-Whitepaper.pdf}{Nvidia’s next generation cuda compute architecture:
  Kepler gk110} (2012).
\newline\urlprefix\url{http://www.nvidia.com/content/PDF/kepler/NVIDIA-Kepler-%
GK110-Architecture-Whitepaper.pdf}

\bibitem{cuda_pg}
NVidia, \href{http://docs.nvidia.com/cuda/cuda-c-programming-guide}{Cuda c
  programming guide} (2013).
\newline\urlprefix\url{http://docs.nvidia.com/cuda/cuda-c-programming-guide}

\bibitem{Sanders:2010:CEI:1891996}
J.~Sanders, E.~Kandrot, CUDA by Example: An Introduction to General-Purpose GPU
  Programming, 1st Edition, Addison-Wesley Professional, 2010.

\bibitem{wilt2013cuda}
N.~Wilt, The CUDA Handbook: A Comprehensive Guide to GPU Programming, Pearson
  Education, 2013.

\end{thebibliography}

%% Authors are advised to submit their bibtex database files. They are
%% requested to list a bibtex style file in the manuscript if they do
%% not want to use elsarticle-num.bst.

%% References without bibTeX database:

% \begin{thebibliography}{00}

%% \bibitem must have the following form:
%%   \bibitem{key}...
%%

% \bibitem{}

% \end{thebibliography}

\end{document}